\documentclass[a4paper,11pt]{article}
\pdfoutput=1 

\usepackage{jinstpub} 

\usepackage{hyperref}
\usepackage{xspace}
\usepackage{booktabs}
\usepackage{siunitx}
\usepackage{verbatim}
\usepackage[table]{xcolor}
\usepackage{subcaption}

\newcommand{\DESYII}{{\mbox{DESY II}}\xspace}
\newcommand{\DIITBF}{{\DESYII Test Beam Facility}\xspace}

\newcommand{\ALICE}{{ALICE}\xspace}
\newcommand{\ALPIDE}{{\textsc{Alpide}}\xspace}

\newcommand{\EUDAQII}{\textsc{EUDAQ2}\xspace}
\newcommand{\EUTELESCOPE}{\textsc{EUTelescope}\xspace}

\newcommand{\MILLEPIDE}{\textsc{Millepide-II}\xspace}

\newcommand{\CORRY}{\textsc{Corryvreckan}\xspace}

\newcommand{\LCIO}{\textsc{LCIO}\xspace}

\newcommand{\EUDET}{\textsc{EUDET}\xspace}

\newcommand{\AIDAII}{\textsc{AIDA-2020}\xspace}

\newcommand{\RunControl}{\texttt{Run Control}\xspace}
\newcommand{\Producer}{\texttt{Producer}\xspace}

\newcommand{\DataConverter}{\texttt{DataConverter}\xspace}

\newcommand{\ADENIUM}{\textsc{ADENIUM}\xspace}

\title{\textsc{ADENIUM } - A demonstrator for a next-generation beam telescope at DESY\note{(c)
All figures and pictures by the author(s) under a \href{https://creativecommons.org/licenses/by/4.0/}{CC BY 4.0} license}}

\author[b, a]{Yi Liu\note{Corresponding author, is now at Zhengzhou University. The work of this paper was mainly done at DESY.} }
\author[c,d]{Changqing Feng}
\author[a]{Ingrid-Maria Gregor}
\author[a]{Adrian Herkert}
\author[a]{Lennart Huth}
\author[a]{Marcel Stanitzki}
\author[c,d]{Yao Teng}
\author[c,d]{Chenfei Yang}

\affiliation[a]{Deutsches Elektronen-Synchrotron DESY, Notkestr. 85, 22607 Hamburg, Germany}
\affiliation[b]{School of Physics and Microelectronics, Zhengzhou University, 450001 Zhengzhou, China}
\affiliation[c]{State Key Laboratory of Particle Detection and Electronics, University of Science and  Technology of China, 230026 Hefei, China}
\affiliation[d]{Department of Modern Physics, University of Science and  Technology of China, 230026 Hefei, China}

\emailAdd{yiliu@zzu.edu.cn}

\abstract{
High-resolution beam telescopes for charged particle tracking are one of the most important and equally demanding infrastructure items at test beam facilities. 
The main purpose of beam telescopes is to provide precise reference track information of beam particles to measure the performance of a device under test (DUT). 
In this report the development of the \ADENIUM beam telescope (\underline{A}LPIDE sensor based \underline{DE}SY \underline{N}ext test beam \underline{I}nstr\underline{um}ent) as a demonstrator and prototype 
for a next-generation beam telescope is presented.
The \ADENIUM beam telescope features up to six pixelated reference planes framed by plastic scintillators for triggering.
ADENIUM is capable of replacing the currently used EUDET-type beam telescopes without impacting existing DUT implementations due to the integration of the telescope DAQ into EUDAQ2.

In this report the concept and design of the \ADENIUM telescope as well as its performance are discussed.
The telescope's pointing resolution is determined in different configurations. For an optimal setup at a momentum of \SI{5.6}{GeV} with an ALPIDE as DUT, a resolution better than \SI{3}{\um} has been extracted.
No rate limitations have been observed at the DESY II test beam.

}
\begin{document}
\maketitle
\flushbottom
\section{Introduction}

Due to the high complexity of detectors for modern particle physics experiments,
it is crucial to demonstrate and validate their performance during all steps of the development and commissioning. 
Test beams with a well defined momentum and particle rate can be used to study the detector performance.
DESY Hamburg operates the \DIITBF~\cite{desytb2018} with three independent beam lines at the \DESYII synchrotron. It is one of the few facilities providing test beams in the \si{\GeV} range worldwide.

A beam telescope is a reference tracking system to reconstruct particle trajectories.
It enables measurements of detector characteristics, such as hit detection efficiency and intrinsic spatial resolution at test beam lines.
The \EUDET-style beam telescopes~\cite{jansen2016} have served as precise reference beam telescopes at the \DIITBF for more than ten years.
They consist of six telescope planes, grouped into two arms, which are framed by plastic scintillators to generate a trigger signal. A trigger-logic-unit~\cite{Baesso_2019} distributes this trigger signal to the telescope and user devices. 
The mechanical support for the telescope is optimized to be easily adjustable for detectors of different sizes. 
The \EUDET-style beam telescopes are about to reach their end of life without sufficient spare sensors and parts in storage, a successor is needed.

The ADENIUM telescope is a key step towards the development of the next generation beam telescope. 
It features a modular DAQ electronics (see section \ref{sec::plane}) and software (see section \ref{sec::daq}) and is based on \ALPIDE~\cite{AglieriRinella:2017lym} sensors, which have been identified as best suited currently available sensors.
ADENIUM can be used interchangeably with EUDET-style telescopes without impacting existing device integration. 
In this report an overview of the \ADENIUM telescope implementation is given and its performance at the \DIITBF is discussed.

\section{The \DIITBF}

DESY operates a test beam facility at the \DESYII synchrotron with three beam lines for detector testing purposes as outlined above.
The \DESYII synchrotron operates in a sinusoidal ramping mode with a frequency of \SI{12.5}{\hertz}. A single electron bunch with a length of \SI{30}{ps} hits a thin carbon fibre serving as primary target in the beam and generates Bremsstrahlung photons.  They are converted back into electron-positron pairs on a metal plate (secondary target). The particles pass a momentum selecting magnet and are delivered to the test beam areas. Users can chose momenta between 1~and~\SI[parse-numbers=false]{6}{GeV} and the particle polarity.

Each of the three test beam areas is equipped with a \EUDET-style beam telescope. A more detailed description of the \DIITBF can be found in reference \cite{desytb2018}.

\section{\ADENIUM Telescope Overview}

The ADENIUM beam telescope consists of six ALPIDE planes 
that are selected for their good spatial resolution and low material to allow precise particle tracking at
low momenta particle beams such as the \DESYII electron beam.  
ADENIUM has to be capable of processing particle rates of a few \SI{10}{kHz} to match the \DESYII test beam rates.

\subsection{Sensor}
Monolithic Active Pixel Sensors (MAPS) are ideal for high resolution tracking telescopes at low momentum beam lines.
 Mimosa26 sensors~\cite{baudot2009,huguo2010} had been chosen for the EUDET-style telescopes, since they provide an
 excellent intrinsic resolution that stems from the fine pixel pitch of \SI{18.4}{\micro m}, combined with diffusion based charge collection in an up to \SI{20}{\micro\meter} thick epitaxial layer.
 Removal of the inactive silicon minimizes the amount of material the particle has to pass.
 
As the long term availability of the Mimosa26 is not clear, the highly available and better performance ALPIDE sensor~\cite{suljic:2658226} is chosen for the ADENIUM telescope.Table~\ref{tab:sensor:parameters}) compares the main
 parameters of the two sensors.
The ALPIDE sensor is implemented in an \SI{180}{nm} CIS process and fabricated on wafers with a \SI{25}{\um} thick high resistivity p-type epitaxial layer on a p-type substrate.   ALPIDE has a slightly larger pitch, but offers a larger active area as well as significantly shorter readout time. 

\begin{table}[htb]
\caption{Main parameters of the Mimosa26~\cite{baudot2009,huguo2010} and \ALPIDE  sensor~\cite{suljic:2658226}.} 
\begin{center}
\begin{tabular}{lcc}
\toprule
 &  {\bf Mimosa} 26 & {\bf \ALPIDE} \\ \hline

Chip sensitive size & 21.2 \si{\milli\meter} $\times$ 10.6  \si{\milli\meter} & 13.8 \si{\milli\meter} $\times$ 29.9 \si{\milli\meter} \\
Chip thickness&  \SIrange{50}{70}{\um} & \SIrange{50}{100}{\um} \\
Pixel pitch & 18.4 \si{\milli\meter} $\times$ 18.4 \si{\micro\meter} & 26.88 \si{\milli\meter} $\times$ 29.24 \si{\micro\meter} \\
Pixel matrix & 1152 $\times$  576  & 512 $\times$  1024\\

Detection efficiency &  \SI{> 99}{\percent} & \SI{> 99}{\percent} \\
Fake-hit rate &\SI{\sim e-4}{pixel^{-1}event^{-1}} & \SI{< e-6}{pixel^{-1}event^{-1}} \\
Typical frame readout time &\SI{115.2}{\us}  &  \SI{10}{\us}\\
\bottomrule
\end{tabular}
\label{tab:sensor:parameters}
\end{center}
\end{table}
Additionally,  the power-efficient design of the chip results in a  negligible heat generation, allowing for passive cooling, which reduces the complexity of the mechanical design significantly compared to the existing MIMOSA26 planes.

\subsection{Telescope Plane\label{sec::plane}}

Each telescope plane can be operated standalone as an independent network node, maximizing the system flexibility. 
An \ADENIUM reference plane, compare Figure~\ref{fig:layer_lab}, consists of three hardware components: the sensor carrier board, the main readout board and a small passive bridge board:
\begin{figure}[ht]
\centering
\includegraphics[angle=0,origin=c,width=0.75\linewidth]{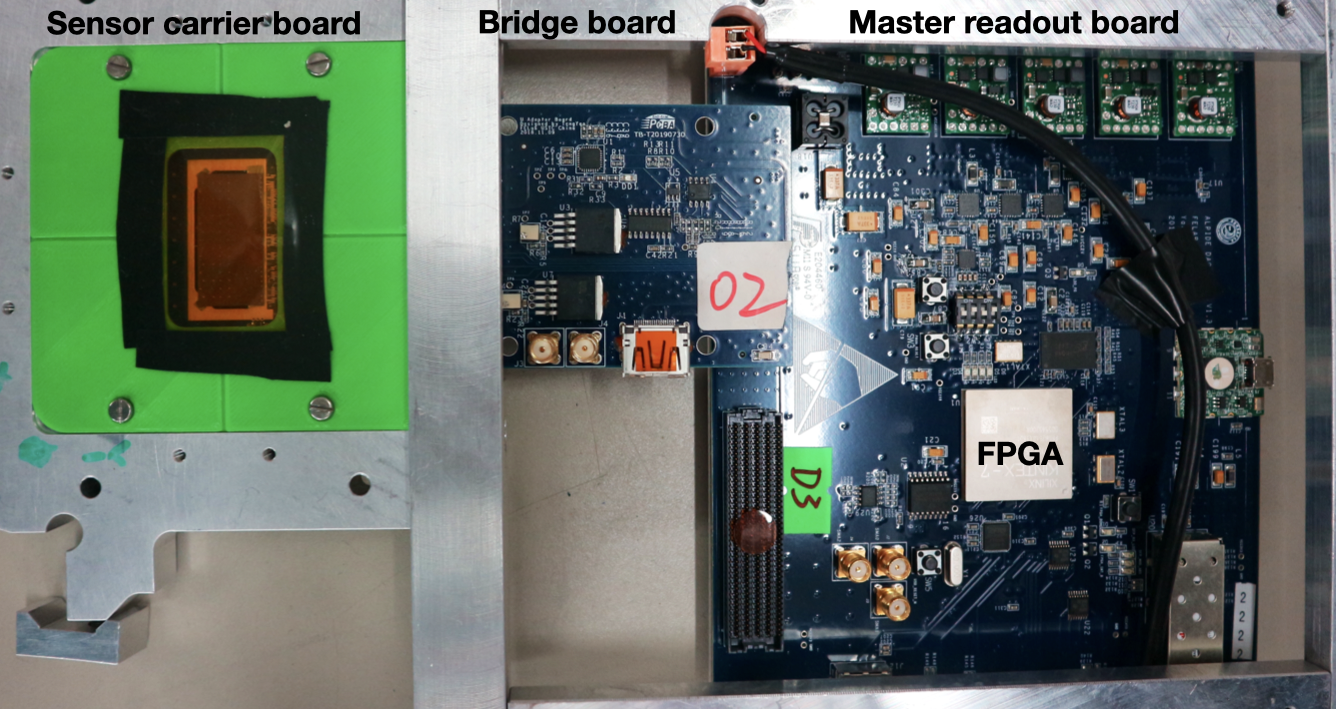} 
\caption{An assembled readout electronic of a telescope plane showing sensor carrier board, bridge board and main readout board.}
\label{fig:layer_lab}
\end{figure}

\begin{itemize}

\item{The {\bf sensor carrier board} is provided by the \ALICE collaboration~\cite{AglieriRinella:2017lym} , with the sensor chip glued and wire bonded to it. 
Below the sensor the PCB has an opening to minimise the material to reduce multiple scattering when particles pass through the sensor. 
Besides the mechanical support for the sensor, the board provides passive electronics for
powering, noise reduction and signal coupling.} 

\item{A {\bf bridge board} is designed, with FMC and PCIe connectors at each end, to bridge between the main readout board and sensor carrier board. 
The bridge board provides the passive electronics for powering and signal coupling, and also carries a CDR (Clock Data Recovery) chip at the up-link data lanes routing from ALPIDE sensor. An optional interface with the \AIDAII Trigger Logic Unit (TLU)~\cite{Baesso_2019} via an HDMI connector is provided on the bridge board.}

\item{The {\bf main readout board} is custom-made at University of Science and Technology of China (USTC). It featues 
a Xilinx Kintex-7 field-programmable gate array (FPGA) chip as the core component~\cite{yang8698804} running a custom firmware to operate the chip and run a server on an ethernet network node. In addition, an optical SFP interface is provided.
The connection to the sensor carrier board is realized via a dedicated Mezzanine Card (FMC) connector.
Power and clock signals for the chip are provided by the main readout board to the sensor over the bridge board. More implementation details of the main read- out board are described in a separate technical article~\cite{teng_9996429} . 
}

\end{itemize}

\subsection{DAQ Software and Data Processing\label{sec::daq}}
\begin{figure}[htb]
  \begin{center}
    \includegraphics[width=1.0\textwidth]{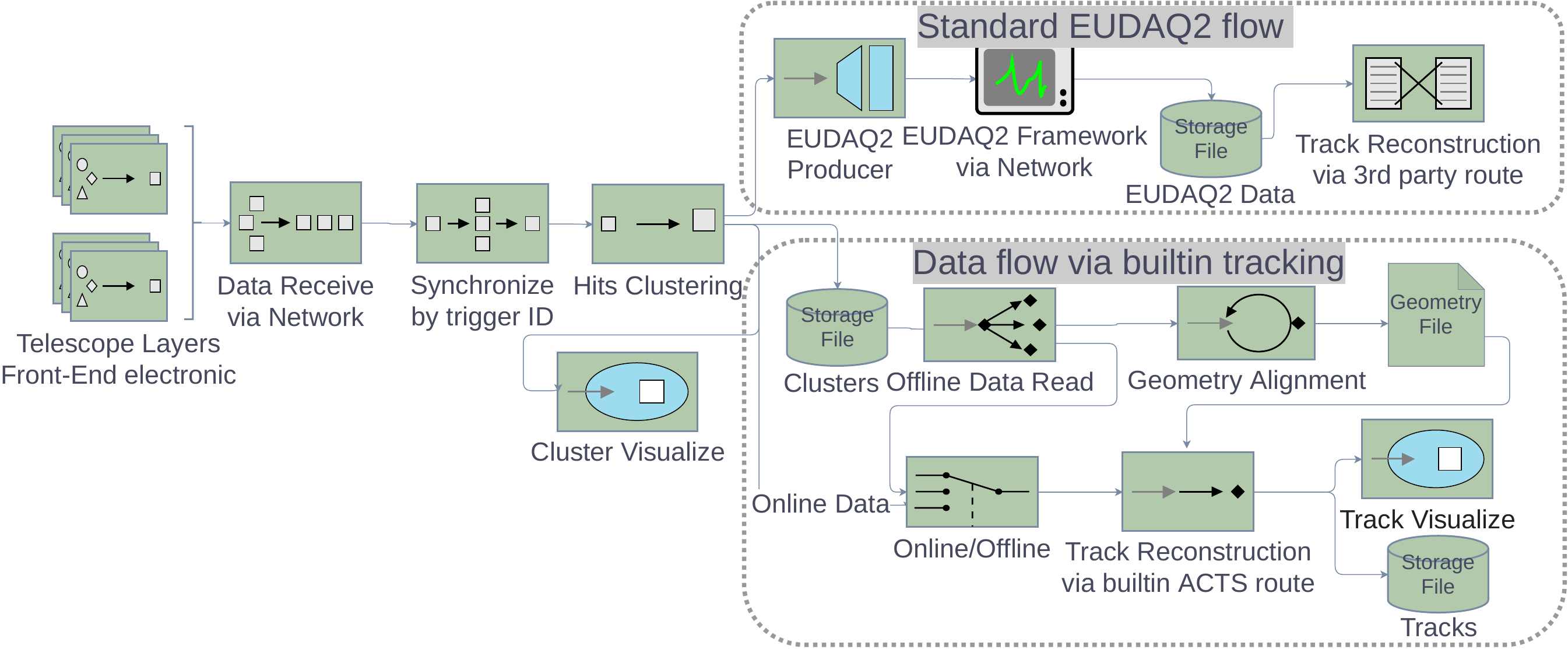}
    \caption{The data flow in \ADENIUM from the individual sensor planes to the the user analysis.}
    \label{fig:adenium_dataflow}
  \end{center}
\end{figure}

The telescope DAQ is realized as a client to all connected telescope planes communicating over a TCP/IP-based interface. The data flow is summarized in Figure~\ref{fig:adenium_dataflow}: the telescope DAQ receives data from all planes, synchronizes it by trigger ID and performs hit clustering. 

A \EUDAQII~\cite{Liu_2019} component, called \Producer, provides the interface between the telescope and EUDAQ2. 
\EUDAQII globally controls all components and logs the data flow including the telescope.
Detector synchronization on hardware level is realised with the \AIDAII trigger-logic-unit \cite{Baesso_2019}.
The central \RunControl of EUDAQ2, with an optional graphical user interface, is able to control the telescope and start/stop its readout via standardized commands processed by the \Producer.
In this concept the reference planes of the telescope are treated the same way as any device under test (DUT)~\cite{Ahlburg:2019jyj}.

According to \EUDAQII specification, a so-called \DataConverter is implemented. A \DataConverter plugin in \EUDAQII framework converts the raw \ADENIUM telescope data to a \EUDAQII native format or the \LCIO format~\cite{Gaede:2003ip}.
Either the built-in tracking discussed below or third party software, like the \CORRY package~\cite{Dannheim:2020jlk}
or the \EUTELESCOPE package~\cite{Bisanz_2020} can be used for the particle trajectory reconstruction.

\subsection{Trigger Implementation}
An \AIDAII Trigger Logic Unit (TLU)~\cite{Baesso_2019} serves as a global trigger distribution system for all connected devices.
A trigger itself is usually generated by a set of scintillator-PMT modules when a particle passes them. 
\ADENIUM uses the {\textit{AIDA-mode-with-id}}. In this mode the TLU sends a trigger, a \SI{40}{\mega\hertz} 
clock, and a trigger ID via an HDMI cable. A busy from the telescope vetos potential additional triggers during readout.

\subsection{Telescope Mechanics}

\begin{figure}[ht]
\centering
\includegraphics[width=0.7\linewidth]{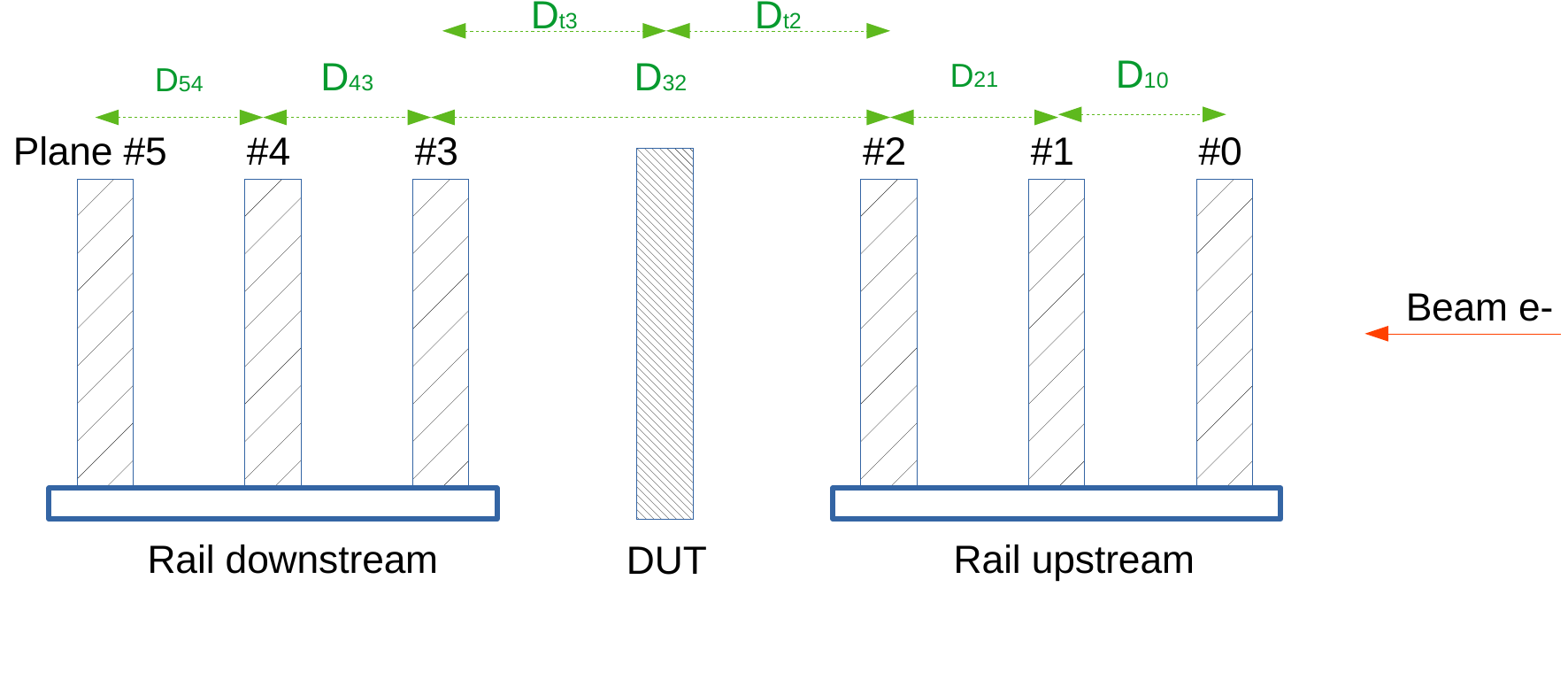}
\caption{Typical arrangement of the  \ADENIUM Telescope planes. 
The beam direction (red line with arrow) is from right to left and the DUT is indicated as a grey box. The telescope planes are 
numbered with consecutive numbers starting from $0$ from right to left. The distances between the planes (green lines with arrows) are denoted by $D_{ij}$. } 

\label{fig:setup}

\end{figure}

\begin{figure}[htb]
  \begin{center}
    \includegraphics[width=0.5\textwidth]{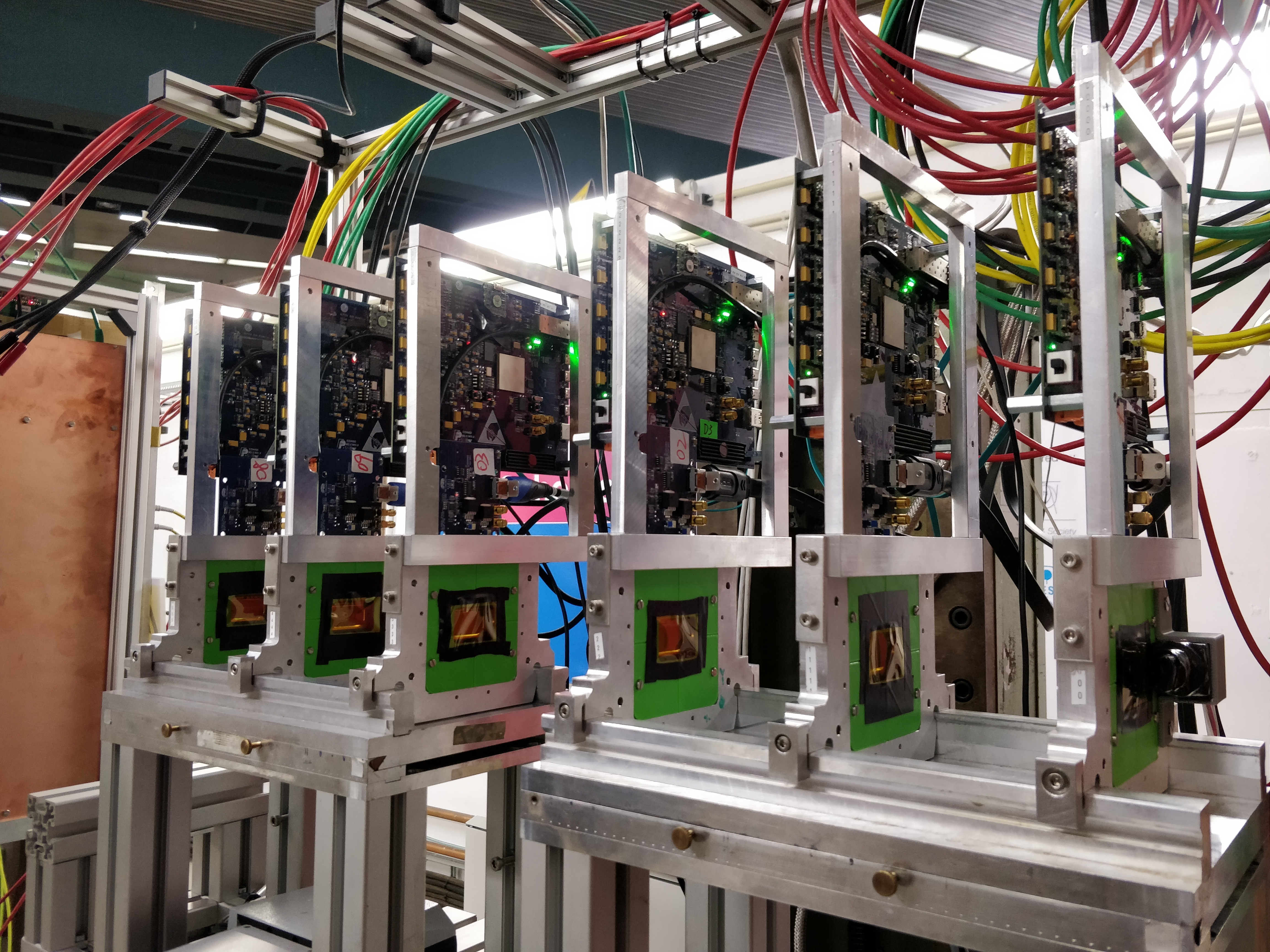}
    \caption{The \ADENIUM  telescope installed at the \DIITBF}
    \label{fig:tele_lab_beam}
  \end{center}
\end{figure}

The telescope mechanics reuses the support structure of the EUDET-type telescopes. Rails for mounting the sensor planes are arranged in two arms parallel to the beam axis (see Figure~\ref{fig:setup}). This design allows for maximum flexibility when integrating user setups. Placing DUTs in the center of the telescope yields the best pointing resolution as the particle trajectories are confined on both sides.

The distance between the two arms is adjustable and can measure up to \SI{50}{\centi\metre} allowing for larger scale DUT setups. The two outermost telescope layers can span a distance of \SI{1.3}{\metre} maximum. A high precision $xy\phi$-stage table can  move the DUT and adjust its position with respect to the active area of \ADENIUM. A photograph of the telescope installed at the \DIITBF is shown in Figure~\ref{fig:tele_lab_beam}.

The sensor carrier board is fixed on an aluminium jig with a cutout that minimises the amount of material in the beam.
The openings are covered by 25 \si{um} thin polyimide sheets on both sides to protect the sensor from dust.
This jig is fastened on the EUDET-style rail system.
The main board is mounted at the top of the jig to an aluminium frame.
Each plane can be moved independently on the rails allowing for a flexible layer arrangement.
No active cooling is implemented in this mechanical support as the overall heat dissipation of the sensor as well as the adjacent electronics is relatively low. The power consumption of ALPIDE is below \SI{40}{mW/cm^2}~\cite{MAGER2016434}. Passive cooling through the jig is sufficient to maintain reasonable temperatures during operation.

\subsection{Integration and synchronization of user detectors}
Correlating the reference trajectories from the telescope with data from DUTs is an essential part of test beam measurements. Hence, it has to be ensured that the telescope and DUT data streams are synchronized. The \AIDAII TLU takes care of this on hardware level by sending a global trigger signal, typically generated by a scintillator coincidence, to all connected detectors. Up to four detectors can be connected to a single \AIDAII TLU. One channel is occupied by ADENIUM, leaving three free slots for user devices. The physical interface must comply with the specification of the AIDA-2020 TLU~\cite{aida2020}. Alongside the trigger signal, the \AIDAII TLU sends a trigger ID, which gets included in the telescope data stream. By including it also in the DUT's data stream the synchronisation gets more robust. The data from both telescope and DUT can be stored in the same file, optionally already sorted by trigger ID.
In addition to the scintillator-based trigger detectors installed by default, user devices can also trigger the AIDA-2020 TLU. It features six trigger inputs with configurable coincidence logic generating the global trigger signal. A precise timestamp of the trigger is recorded by the TLU and can be added to the data stream. 

An optional full integration of the user device in \EUDAQII allows for synchronized configuring,
starting and stopping of the telescope's and DUT's DAQs.
Data can be be saved in the \EUDAQII native data format, where an event package is the fundamental storage structure, which contains the trigger ID, timestamp, telescope data, and DUT hit data belonging to a single trigger. 
The size of an event package per trigger varies and depends on the telescope's occupancy and the DUT data defined by the user. Typically, at the \DIITBF less than three clusters are recorded per telescope plane and trigger. In this case, without a DUT, the size of the event package is about 1~kByte.

\subsection{Built-in Track Reconstruction Module\label{sec:Teleintr:track}}

The ADENIUM DAQ software features built-in track reconstruction based on
a combinatorial Kalman Filter (CKF) tracking algorithm developed as a part of the ACTS project~\cite{ai2021common}. The latter provides a toolkit for track reconstruction in a generic, 
framework- and experiment-independent software package. The fast CKF algorithm of the built-in tracking module can run online and offline (see Section~\ref{sec:Performance}). In online mode each trigger event is processed immediately. The offline mode can be used with a refined software alignment of the telescope planes.

\begin{figure}[htbp]
  \begin{center}
    \includegraphics[width=0.9\textwidth]{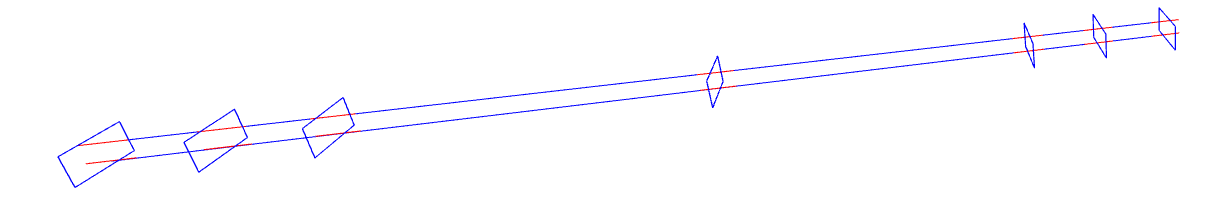}
    \caption{A 3D visualization of reconstructed double trajectories belonging to a single trigger (beam energy at \SI{2.0}{GeV}) by the DAQ software. 
    The blue rectangles are the sensitive areas of the telescope. The red segmented lines crossing the sensitive areas denote the measured hit clusters by the telescope planes. 
    The telescope planes are placed as in Figure~\ref{fig:setup}. The distances ($D_{ij}$) between the planes are configured to be
    $D_{10}=D_{21}=D_{43}=D_{54}=38~mm$ and $D_{t2}=D_{t3}=D_{32}/2=150~mm$.}
    \label{fig:dual_tarjs_display}
  \end{center}
\end{figure}

Since multiple beam particles could potentially pass through the telescope in the same trigger window, the CKF algorithm is configured to take all clusters in the first telescope plane as seed 
of potential track candidates. For each seed, the CKF algorithm performs the forward track propagation into the next telescope plane. Clusters in the following telescope plane are scored using the predicted $\chi^2$ obtained by comparing the Kalman filter prediction and the measured cluster center with the resolution of the cluster taken into account~\cite{Fruhwirth:1987fm}. The cluster with the highest score within a search window is added to the track. 
The track parameters are then updated for next forward track propagation. The CKF algorithm is tolerant to imperfect trajectory with missing clusters, due to either sensor inefficiency or to the trajectory being partially outside of the sensitive region (misalignment). A backward track parameter smoothing is performed afterwards to further improve the precision.
The reconstructed particle trajectories can be stored in the same file alongside the original telescope pixel hit and cluster data.
The telescope DAQ software has built-in 3D graphic window, implemented using OpenGL, to visualise the reconstructed trajectories in real-time. 
The graphical display shows only the last reconstructed event.
Figure ~\ref{fig:dual_tarjs_display} shows two reconstructed trajectories which belongs to a single trigger 
with a beam energy of \SI{2.0}{GeV} and high beam rate.

To not deteriorate the precision of the reconstructed tracks, the geometry of the telescope, i.e.~the position and orientation of the telescope planes, must be precisely known.
The built-in tracking module provides tools to perform a track based alignment using the \MILLEPIDE~\cite{millepede2:2006} package, which can significantly improve the precision 
of the telescope geometry after several iterative runs. For the time being, the alignment procedure is standalone and supposed to be executed prior to the track reconstruction. 
The generated geometry file with misalignment corrected is then fed into the online track reconstruction.

\section{Performance Studies\label{sec:Performance}}
A series of measurements was performed to verify the functionality of the \ADENIUM telescope 
and to characterize its performance in terms of timing and rate capability, noise, hit detection efficiency, cluster sizes and pointing resolution. The telescope arrangement shown in Figure~\ref{fig:setup} was used. An additional \ADENIUM telescope plane is installed as DUT for the track resolution study. 
Note that all sensors are currently configured with identical settings, hence the results can be further optimized with sensor-specific settings, e.g.~ bias voltage, threshold reference, sampling timing and so on.  

\subsection{Timing and Trigger Rate}
\label{subsec:trigger}
While the TLU is capable to trigger up to a rate of \SI{10}{\mega\hertz} and provides a precise time stamp with \SI{781}{\pico\second} binning,  the overall 
trigger rate of the telescope is limited by its sensors. Their analog circuit has a peaking time of about \SI{2}{\micro\second}~\cite{AGLIERIRINELLA2017583}. A readout time of ca. \SI{10}{\micro\second}
is chosen accordingly to ensure that the hit that corresponds to the trigger is read out. Therefore, the maximum processable trigger rate is about \SI{100}{\kilo\hertz} - far more than can be achieved with the \DESYII test beams. At the \DIITBF a maximum trigger rate of \SI{40}{\kilo\hertz} was measured and no rate dependent limitations of \ADENIUM were observed. Note that multiple hits can be recorded per trigger and the telescope is not able to resolve them in time, since no on-chip time stamping is provided. Such ambiguities can be resolved by incorporating one additional tracking plane consisting of a detector with high spatial and time resolution \cite{Obermann_2014}.  

\subsection{Noise and Hit Detection Efficiency}
\label{subsec:noise}
Due to variations in CMOS processes, the sensors have slightly different noise levels at
identical configurations. A pixel scoring approach has been used to classify the pixel quality. 

\begin{table}[htbp]
\caption{The fraction of pixels per plane belonging to the three noise-occupancy categories}
\begin{center}
\begin{tabular}{c c | c c c c c c c} 
Grade & Noise Occupancy & Plane \#0 & Plane \#1 & Plane \#2 & Plane \#3 & Plane \#4 & Plane \#5 \\
\toprule
A & $<10^{-6}$          &99.975\% &99.998\% &99.991\% &99.973\% &99.967\% &99.988\%  \\
B & $10^{-6}<N<10^{-3}$ & 0.022\% & 0.002\% & 0.008\% & 0.024\% & 0.029\% & 0.011\% \\
C & $>10^{-3}$          & 0.003\% & 0.000\% & 0.001\% & 0.003\% & 0.004\% & 0.001\% \\
\bottomrule
\end{tabular}
\label{tab:pixel_grade}
\end{center}
\end{table}
With the beam turned off, the TLU issues a million triggers to all telescope planes to sample all pixels. 
Ignoring the negligible impact of the natural background, the noise occupancy of all individual pixels can be 
measured by counting the noise hits per pixel. Pixels are then categorized depending on their noise occupancy 
(see Table ~\ref{tab:pixel_grade}). The grade B and grade C pixels can be added to a configuration file to disable them.

\begin{table}[htbp]

\caption{The hit efficiencies for each telescope plane at two beam energies}
\begin{center}
\resizebox{1.0\textwidth}{!}{
\begin{tabular}{c| c c c c c c c} 
Energy & Plane \#0 & Plane \#1 & Plane \#2 & Plane \#3 & Plane \#4 & Plane \#5 \\
\toprule
5.6 \si{GeV} & (99.92$\pm$0.01)\% & (99.76$\pm$0.02)\% & (99.87$\pm$0.01)\% & (99.98$\pm$0.00)\% &  (99.92$\pm$0.01)\% & (99.48$\pm$0.02)\% \\
2.0 \si{GeV} & (99.90$\pm$0.01)\% &  (99.56$\pm$0.02)\% &  (99.82$\pm$0.02)\% &  (99.95$\pm$0.01)\% &  (99.84$\pm$0.01)\% &  (98.79$\pm$0.04)\% \\
\bottomrule
\end{tabular}
}
\label{tab:efficiency_per_layer}
\end{center}
\end{table}

The hit detection efficiency is measured using the ratio of reconstructed tracks with no hit on the plane under test to all reconstructed tracks. 
"No hit" is defined here as no cluster being found with the distance less than \SI{0.5}{\mm} to the expected intersection of the track, taking the effect of multiple scattering into account when matching the hit to the tracks. The hit efficiency of the different telescope planes is shown in Table~\ref{tab:efficiency_per_layer}.

\subsection{Hit Cluster}
\label{subsec:cluster}
Due to charge sharing, several neighbouring pixels may respond to a single particle hit at the same time. The number of pixels belonging to a hit cluster, 
i.e.~the cluster size, depends on the incident angle of the particle as well as its hit position within the pixel.

\begin{figure}[htbp]
\begin{subfigure}[t]{0.48\textwidth}
\includegraphics[width=0.99\linewidth]{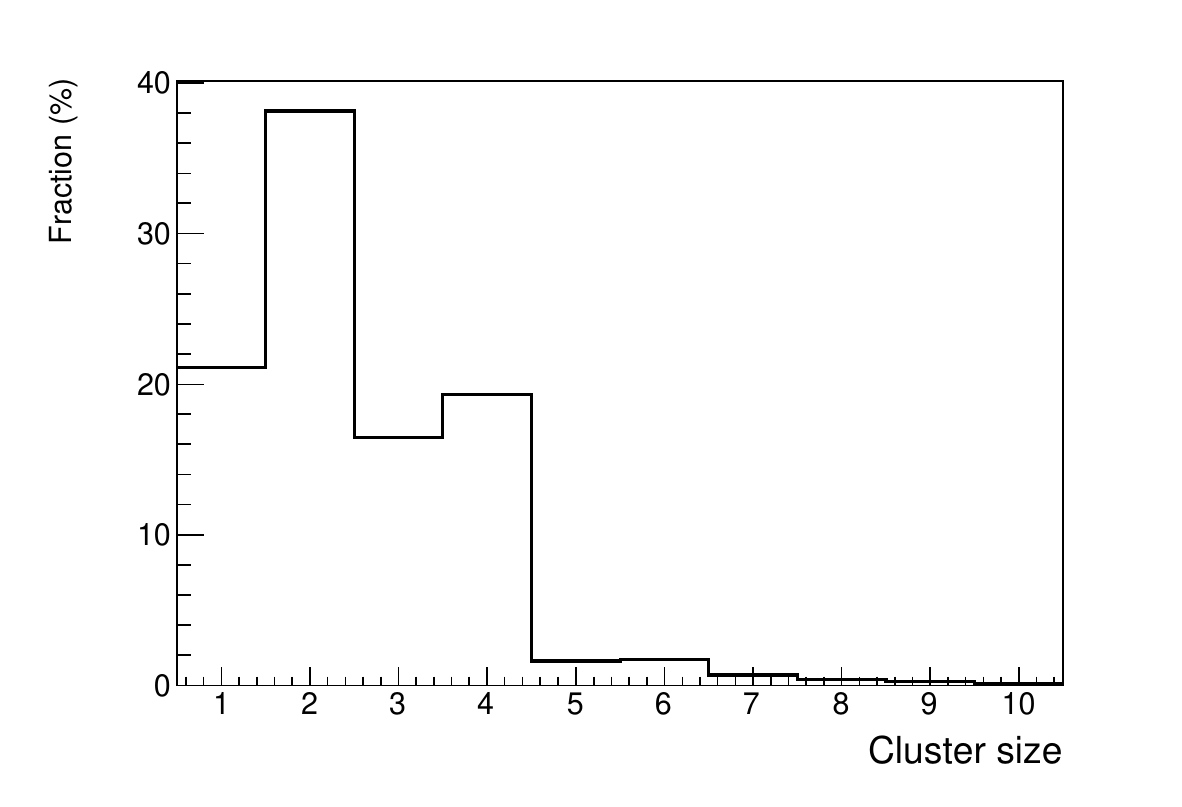} 
\caption{}
\label{fig:clusterSize_trajHit_DetN1_layer}
\end{subfigure}
\begin{subfigure}[t]{0.48\textwidth}
\includegraphics[width=0.99\linewidth]{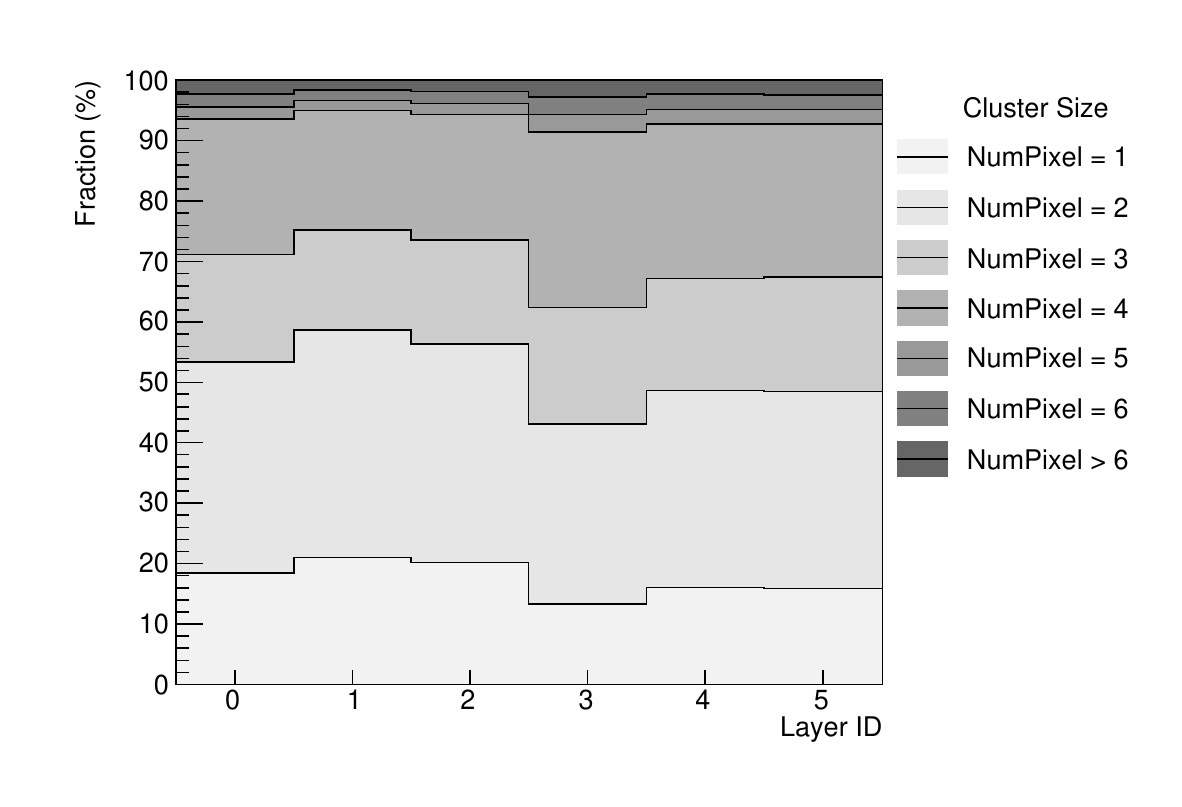}
\caption{}
\label{fig:clusterSize_trajHit_all_layer}
\end{subfigure}
\caption{(a) Fraction of cluster sizes for one telescope plane (layer ID = 1). (b) Fraction of clusters with sizes 1 to 6 for the six telescope planes. }
\end{figure}

\begin{table}[htbp]

\caption{The mean cluster size and its projected sizes in both dimensions}
\begin{center}
\resizebox{1.0\textwidth}{!}{
\begin{tabular}{c| c c c c c c c} 
 & Plane \#0 & Plane \#1 & Plane \#2 & Plane \#3 & Plane \#4 & Plane \#5  & mean of all planes\\
\toprule
mean number of pixels per cluster& 2.71 & 2.56 & 2.62 & 2.99 & 2.81 & 2.86 & 2.76\\
x dimension (pixel pitch \SI{29.24}{\micro\metre}) & 1.65 &  1.60 &  1.62 &  1.74 &   1.68 &  1.70  &  1.67 \\
y dimension (pixel pitch \SI{26.88}{\micro\metre})&  1.73 &  1.68 &  1.70 &  1.83 &   1.77 &  1.79 &  1.75  \\
\bottomrule
\end{tabular}
}
\label{tab:pixel_size_per_layer}
\end{center}
\end{table}

Figure~\ref{fig:clusterSize_trajHit_all_layer} shows the percentage of clusters with different sizes for each telescope plane. Table~\ref{tab:pixel_size_per_layer} lists the mean of cluster sizes as well as projected sizes in x and y dimensions. The x and y dimensions have pixel pitch of \SI{29.24}{\micro\metre} and \SI{26.88}{\micro\metre}, respectively.
The mean number of pixels is 2.76 pixels per cluster and varies among different sensors.
This can be further optimized and adjusted to a uniform value for all sensors, e.g. via the comparator thresholds.
For single-pixel clusters, the intrinsic resolution of the sensor can be estimated as \SI{8.44}{\micro\meter} and \SI{7.76}{\micro\metre} in the 
$x$ and $y$ dimension, respectively, using the formula: \(d/\sqrt{12}\), where $d$ is the pixel pitch. Charge sharing can improve this resolution when the cluster center is used as the measured hit position.

The average cluster size of all sensors is independent of beam energies ranging from \SI{2.0}{\GeV} to \SI{5.6}{\GeV}, hence the intrinsic resolution at the \DESYII beam lines is constant, as expected.

\subsection{Track Interpolation}
\label{subsec:trackext}
\begin{figure}[htbp]
\begin{subfigure}[t]{0.48\textwidth}
\includegraphics[width=0.99\linewidth]{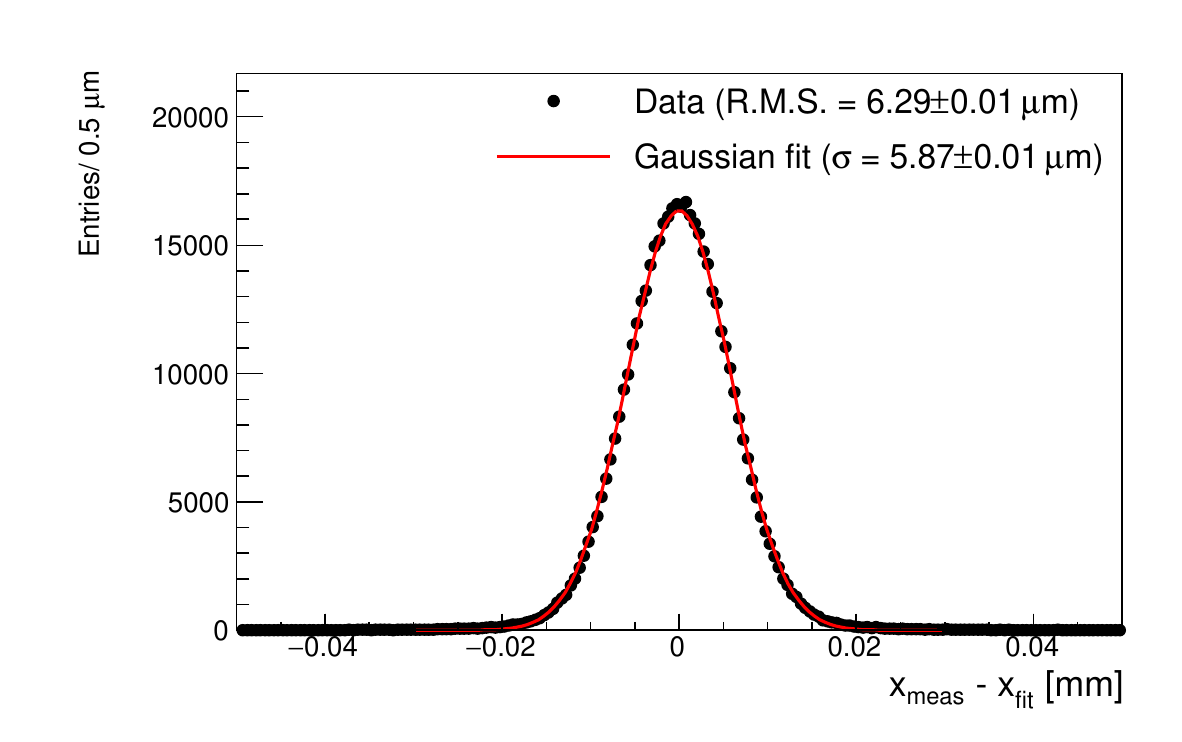} 
 \caption{$x$ direction (pixel pitch is \SI{29.24}{\micro\metre})}
\label{fig:res_geo7_gev5p6_x}
\end{subfigure}
\begin{subfigure}[t]{0.48\textwidth}
\includegraphics[width=0.99\linewidth]{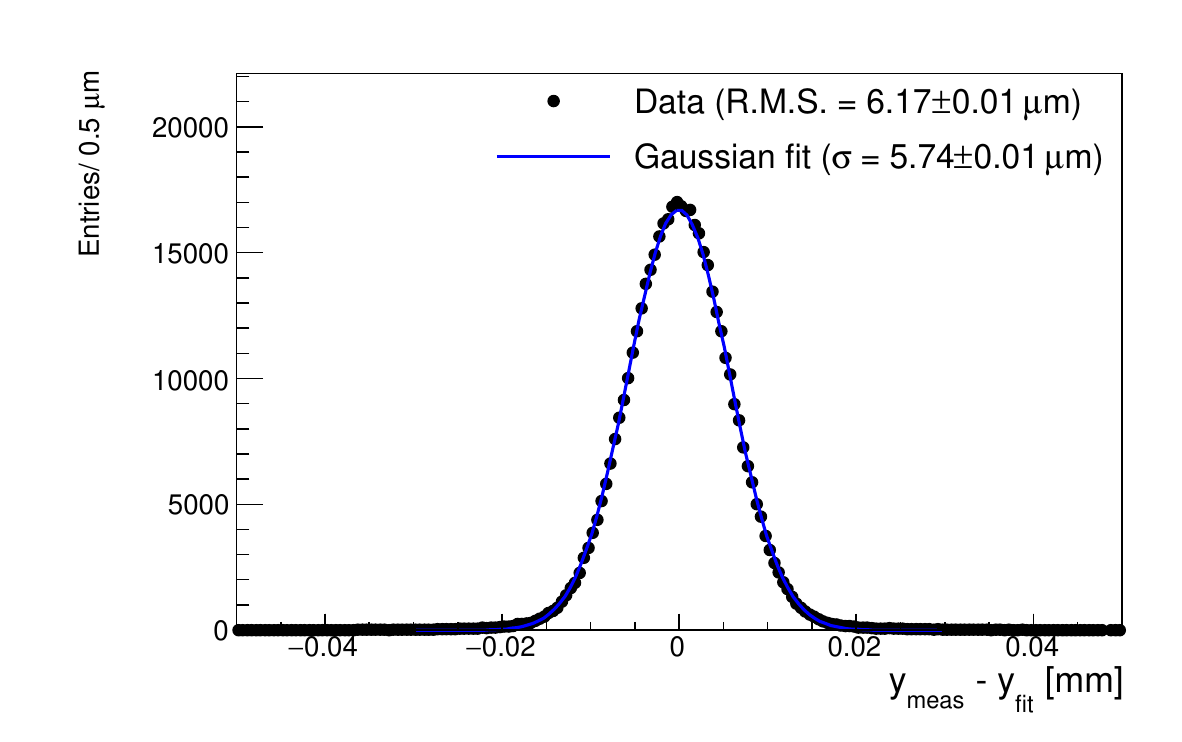}
\caption{$y$ direction (pixel pitch is \SI{26.88}{\micro\metre})}
\label{fig:res_geo7_gev5p6_y}
\end{subfigure}
\caption{Distributions of residual in the (a)$x$ and (b)$y$ dimension of the DUT plane for a sample of 500k electrons at \SI{5.6}{GeV}. 
The six telescope planes with additional telescope plane taken as the DUT are grouped as shown in Figure~\ref{fig:setup}. The distances ($D_{ij}$) between the planes are configured to be $D_{10}=D_{21}=D_{43}=D_{54}=D_{t2}=D_{t3}=$ \SI{38}{\milli\meter}. The black dots are data, and the red (blue) lines are Gaussian fits to the data.}
\label{fig:res_geo7_gev5p6_t3}
\end{figure}

With a properly aligned detector geometry, the built-in analysis module of the DAQ software (see Section~\ref{sec:Teleintr:track})
in the stand-alone mode is able to reconstruct the particle trajectories
and their intersection points with DUT planes online. 
At each propagation step of the CKF algorithm, the predicted $\chi^2$ is expected to follow a distribution with two degrees of freedom for the each telescope plane. A cut of 13.8 is imposed on the predicted $\chi^2$ when the clusters are associated to the track in the CKF algorithm.  This corresponds to accepting hits stemming from the correct  particle with a probability of 99.9\%.
The interpolated position of a reconstructed track on the DUT plane can be compared to the measured DUT hit.
Figure~\ref{fig:res_geo7_gev5p6_t3} shows these distributions of unbiased residuals, i.e.~excluding the DUT hits from the track fit, for a sample of 500k electrons at \SI{5.6}{GeV} with the telescope setup as
shown in Figure~\ref{fig:dual_tarjs_display}. Two Gaussian fits to 
the unbiased residuals show a width of  $\sigma=5.87\pm0.01$~\si{\micro\meter} and $\sigma=5.74\pm0.01$~\si{\micro\meter} in the $x$ and $y$ dimension of the DUT plane, respectively.

\begin{figure}[ht]

\begin{center}
\includegraphics[width=0.6\textwidth]{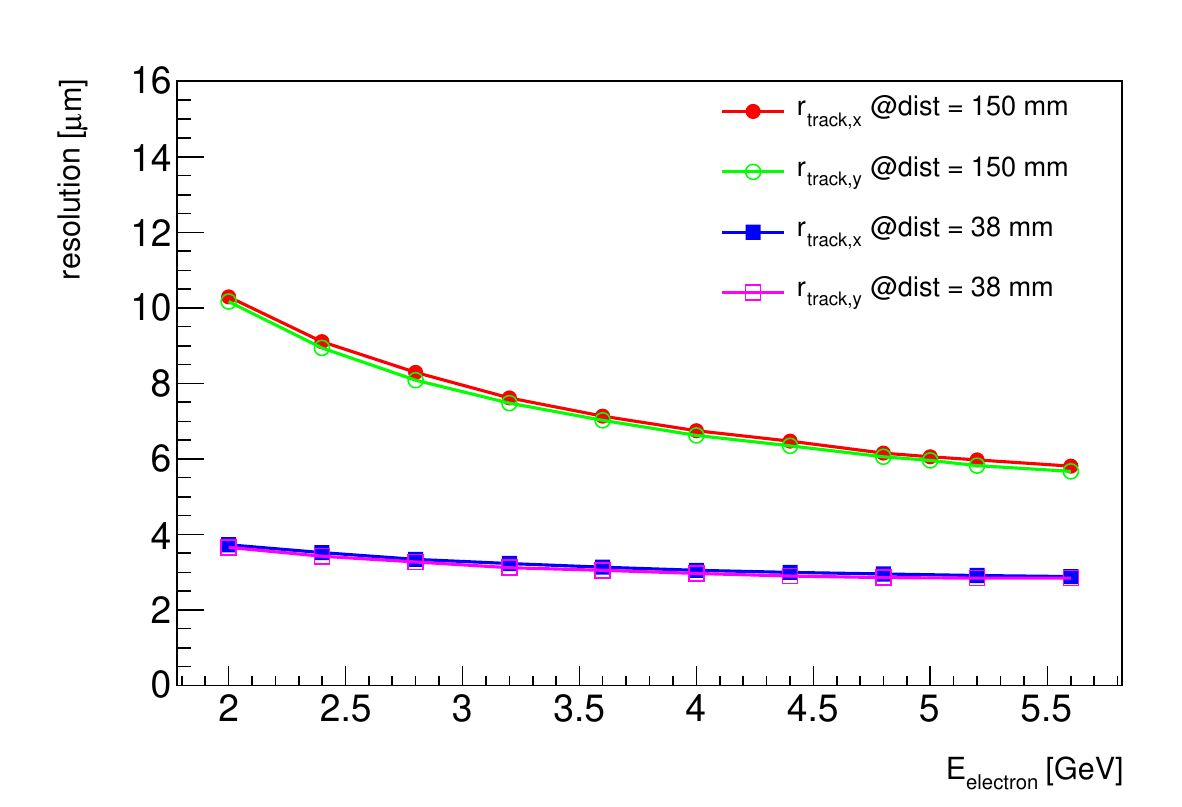} 
\end{center}
\caption{Resolution of the telescope at  an additional telescope plane acting as DUT with distances
among the planes of $D_{10}=D_{21}=D_{43}=D_{54}=$ \SI{38}{\milli\meter} and $D_{t2}=D_{t3}=D_{32}/2$. The different markers represent different distances to the innermost telescope planes ($D_{t2},D_{t3}$). The slight difference in resolution along x/y is due to the difference in pixel pitch.}

\label{fig:resScan}

\end{figure}

The resolution of the telescope at DUT position can be evaluated based on the measured biased residual and DUT intrinsic resolution with the following formula:
\begin{equation}
r_{track} = \sqrt{\sigma_{unbias}^2 - r_{dut}^2}.
\end{equation}

The intrinsic resolution of DUT plane is estimated using the geometric mean ~\cite{CARNEGIE2005372} of the biased and the unbiased residuals:
\begin{equation}
r_{dut} = \sqrt{\sigma_{unbias} \cdot \sigma_{bias}}.
\end{equation}
where the biased residual comes from the interpolated position of a reconstructed track with the inclusion of DUT hit position compared to the DUT hit position. The resolution of the telescope for different spacing as a function of the momentum is shown in Figure~\ref{fig:resScan}. 
As expected, the resolution improves with increasing momentum since the impact of multiple scattering is reduced.
The dependency of the tracking precision on the distances from the DUT to innermost telescope planes can be seen as well, i.e.~smaller distances result in a higher precision on the DUT. 
This can be explained by less air that has to be passed as well as smaller uncertainties on the propagation. In the optimal setup, a resolution of $r_{track,x} = \SI{2.89}{\um}$ and $r_{track,y}  =\SI{2.84}{\um}$ at \SI{5.6}{\GeV} could be determined. 

\section{Summary and Outlook}
A demonstrator of a high resolution telescope, as upgrade and drop-in replacement 
of the \EUDET-style telescopes, was developed and tested. It reaches pointing resolution below \SI{3}{\micro\meter} and shows no performance limitations at the maximum trigger rates that can be reached at the DESY II test beams (ca. \SI{40}{\kilo\hertz}). 
It is compatible with the existing \EUDET-style infrastructure, so 
users can switch to the new telescope easily without additional integration effort. 
The compact readout electronics allow a flexible arrangement of the telescope 
planes. 
The demonstrated performance (see in section \ref{sec:Performance}, i.e.~trigger rate, tracking resolution, sensor noise and detection efficiency) justifies the choice of the sensor and the DAQ layout.
An upgrade of all \EUDET-style pixel beam telescopes in the near future will be based on the presented system.  
The next development for a long-term version of the \ADENIUM telescope is on-going. 
A more compact readout electronic based on a Xilinx Zynq SOC has been designed and manufactured.
It will allow the implementation of more advanced features including running the DAQ software on the front-end Zynq SOC itself.

\section*{Acknowledgment}  

The authors would like to acknowledge Deutsches Elektronen-Synchrotron DESY and Helmholtz Gesellschaft Funding for all the support.
The measurements leading to these results have been performed at the test beam facility at DESY Hamburg (Germany), a member of the Helmholtz Association.
The authors would like to thank the technical team at the \DESYII accelerator and the \DIITBF  for the smooth operation of the test beam
and the support during the test beam campaign.
The readout electronics of this work was supported in part by the National Natural Science Foundation of China 
under Grant 11922510 and Grant 11773027, and in part by the National Key Technologies Research and Development Program under Grant 2016YFE0100900.
\bibliography{alpidebibfile}

\end{document}